\documentclass[preprint]{vgtc}               

\graphicspath{{figures/}{pictures/}{images/}{./}} 

\usepackage{times}                     

\usepackage{tabu}                      
\usepackage{booktabs}                  
\usepackage{lipsum}                    
\usepackage{mwe}                       

\usepackage{soul} 
\usepackage{color}
\usepackage{colortbl}
\usepackage{balance}

\usepackage{mathptmx}                  
\usepackage{comment}
\usepackage{tabularx}

\newif\ifhighlightchanges

\newcommand{\newmarker}[1]{%
\ifhighlightchanges
\textcolor{blue}{#1}%
\else
#1%
\fi
}

\newcommand{\deletemarker}[1]{%
\ifhighlightchanges
\textcolor{blue}{\st{#1}}%
\else
\fi
}

\highlightchangesfalse
\definecolor{mynewmarker}{rgb}{0,0,0}

\onlineid{6688}

\vgtccategory{Research}

\vgtcinsertpkg

\title{Accented Character Entry Using Physical Keyboards in Virtual Reality}

\def\thanksANote#1{%
  \footnotemarkANote%
  \protected@xdef\@thanks{\@thanks%
        \protect\footnotetextANote[\the \c@footnoteANote]{#1}}%
}

\author{\textsuperscript{\$}Snehanjali Kalamkar\thanks{e-mail: snehanjali.kalamkar@hs-coburg.de } \\ %
        \scriptsize Coburg University  %
\and \textsuperscript{\$}Verena Biener\thanks{e-mail: verena.biener@visus.uni-stuttgart.de}\\ %
     \scriptsize University of Stuttgart %
\and Daniel Pauls\\ %
     \scriptsize Coburg University
\and Leon Lindlein\\ %
     \scriptsize Coburg University
\and Morteza Izadifar\\ %
     \scriptsize Coburg University
\and Per Ola Kristensson\\ %
     \scriptsize University of Cambridge
\and Jens Grubert\thanks{e-mail: jens.grubert@hs-coburg.de \\ \textsuperscript{\$}Snehanjali Kalamkar and Verena Biener contributed equally to this work.}\\ %
     \scriptsize Coburg University %
     }

\teaser{
  \centering
  \includegraphics[width=\linewidth]{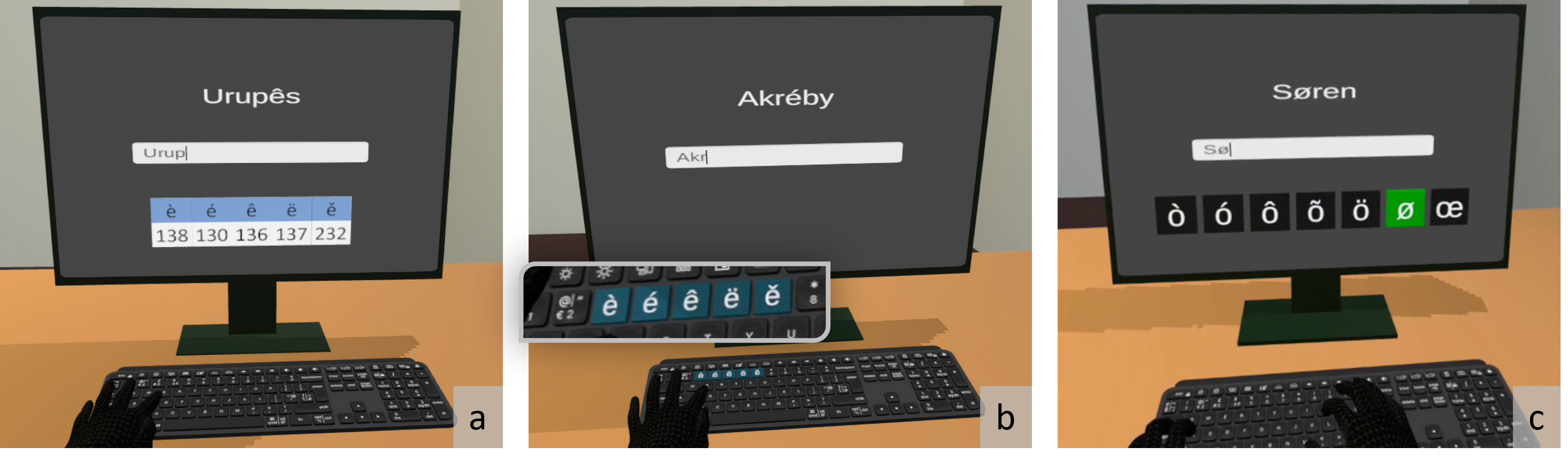}
  \caption{%
  Techniques for entering accented characters in VR: a) \textsc{baseline} technique which displays numeric codes for relevant accented characters after pressing a modifier key (Alt) and a base character and then requires the user to enter the respective numeric code; b) \textsc{reconviguration} technique as presented by Schneider et al. \cite{schneider2019reconviguration} which overlays the accented characters onto the keyboard when pressing a modifier key (Alt) and a base character; c) \textsc{multimodal} technique which presents the accented characters on the virtual screen while pressing a modifier key (Alt) and a base character and enables the user to select one via eye-gaze and confirm by key press.
  }
  \label{fig:teaser}
}

\abstract{
Research on text entry in Virtual Reality (VR) has gained popularity but the efficient entry of accented characters, characters with diacritical marks, in VR remains underexplored.
Entering accented characters is supported on most capacitive touch keyboards through a long press on a base character and a subsequent selection of the accented character. 
However, entering those characters on physical keyboards is still challenging, as they require a recall and an entry of respective numeric codes.
To address this issue this paper investigates three techniques to support accented character entry on physical keyboards in VR. Specifically, we compare a context-aware numeric code technique that does not require users to recall a code, a key-press-only condition in which the accented characters are dynamically remapped to physical keys next to a base character, and a multimodal technique, in which eye gaze is used to select the accented version of a base character previously selected by key-press on the keyboard.
The results from our user study ($n=18$) reveal that both the key-press-only and the multimodal technique outperform the baseline technique in terms of text entry speed. 

} 


\begin{document}


\firstsection{Introduction}

\maketitle


Text entry is a fundamental activity in Virtual Reality (VR), and as a consequence researchers have explored a wide range of techniques~\cite{dube2019text}.
Besides using physical~\cite{knierim2018physical, grubert2018effects} and virtual keyboards~\cite{dudley2023evaluating}, researchers have explored gestures~\cite{bowman2001pinch,shen2023fast}, controllers~\cite{speicher2018selection}, head tracking~\cite{yu2017tap} and eye gaze~\cite{rajanna2018gaze,hu2024skimr}.

However, while there has been extensive research efforts directed towards general text entry, there has been limited attention on how to enter accented characters, which has been previously raised as a text entry challenge, even outside of VR~\cite{kristensson2009five}.
On physical keyboards, special numeric codes need to be recalled to type many accented characters. For example, Microsoft Windows enables users to input accented characters through pre-defined numeric codes (such as Alt+143 to obtain Å), while Linux uses Unicode. 
Linux also provides the ability to combine a sequence of characters to enter a specific accented character using a so-called Compose key. For example, pressing the Compose key followed by the characters A and an apostrophe (') to obtain Á, and similarly, the apostrophe can be applied to the characters E, I, O, and U to obtain É, Í, Ó, and Ú, respectively.
These techniques either require users to recall and then enter the respective numeric codes or Unicode, or they recall custom sequences, such as with the Compose Key.
Such techniques become cumbersome very quickly when the users' need for accented characters increases, such as when they are writing for an international audience or when typing words with diacritical marks, such as entering ``naïve'' instead of ``naive''.

Prior work has suggested that VR can be used to enhance the capabilities of physical keyboards.
For example, Schneider et al.~\cite{schneider2019reconviguration} proposed an approach they dubbed `reconviguration', which means that the keys on a physical keyboard are given new visual appearances and functions in VR depending on the application. Reconviguration allows changing the keyboard layout to facilitate the input of accented characters and foreign scripts, or can be used to completely change the layout of the keyboard by combining multiple keys into one larger button~\cite{schneider2019reconviguration}.
However, while prior work~\cite{schneider2019reconviguration} proposed a technique for accented character entry in VR, this technique was not formally evaluated.

Inspired by, and extending, this prior work~\cite{schneider2019reconviguration}, we propose three techniques for supporting accented character entry using physical keyboards in VR. 
First, as the \textsc{baseline}, we introduce a technique inspired by the Microsoft Windows numeric code technique. We present a context-aware numeric code method that displays relevant codes to the user, and therefore does not require the user to recall the specific codes for entering a desired accented character. 
Second, we adopt the \textsc{reconviguration} technique proposed by Schneider et al. ~\cite{schneider2019reconviguration} that changes the keyboard layout to display accented characters. Third, we introduce a \textsc{multimodal} technique that combines eye gaze pointing with key presses to allow the user to select accented characters. 
A controlled user study that compares the performance of these three techniques showed that the \textsc{reconviguration} and \textsc{multimodal} techniques enabled significantly faster text entry than \textsc{baseline}. Specifically \textsc{reconviguration} was 33.6\% faster than \textsc{baseline} for overall words and 72.7\% faster when entering accented characters. Similarly, \textsc{multimodal} was 25.17\% faster than \textsc{baseline} for overall words and was 76.4\% faster for entering accented characters. 
In addition, our findings showed a significantly higher (42\%) usability of the \textsc{reconviguration} technique than \textsc{baseline}.
Finally, \textsc{reconviguration} and \textsc{multimodal} were preferred over the other techniques by 44.4\% of the participants each.


\section{Related Work}
This section provides an overview of recent research on text entry techniques for VR, especially using physical keyboards, and summarizes prior work on typing special or accented characters. In addition, it provides a short overview of prior work in the area of multimodal interactions that involve eye-gaze.

\subsection{Text Entry Techniques in VR}
In the last years, there has been extensive research on how to enter text in VR and a review of such work has been provided by Dube et al. \cite{dube2019text}.
Previous research has explored different modalities for typing in VR.
One of them is controllers \cite{boletsis2019controller, chen2019exploring, wan2024design, speicher2018selection}, which have been found to be more usable than text entry through a touch device \cite{chen2019exploring} and outperforms other methods such as head pointing or freehand input \cite{speicher2018selection}.
Others proposed to use gestures using pinch gloves \cite{bowman2001pinch}, by swiping a cursor over a virtual keyboard through the rotations of the hand \cite{gupta2019rotoswype}, or by typing on a virtual keyboard on a surface or mid-air \cite{dudley2019performance}.
This could also be combined with statistical decoders to predict the intended words and therefore allow for faster and more accurate text input \cite{dudley2018fast}.
It has also been shown to be feasible to type on the users' palm \cite{wang2015palmtype}.
Another approach for text entry in VR is speech recognition \cite{adhikary2021text, pick2016swifter} which has shown to be faster than mid-air typing \cite{adhikary2021text}.

As the VR system inherently tracks the users' head movement, head-gaze has also been proposed as a modality for text entry \cite{yu2017tap}.
To avoid dwell time to confirm the selection of certain characters Lu et al. \cite{lu2020exploration} explored blinking, or neck movements for confirmation.
Another approach to avoid dwell time was proposed by Xu et al. \cite{xu2019ringtext} by arranging the letters in a circle.
In extension to head-gaze, there are also multiple techniques that employ eye-gaze for text entry.
This has been explored also outside of VR, for example, in the context of handicapped people \cite{majaranta2002twenty}. 
There have also been multiple approaches on how to avoid dwell time, such as adjusting it depending on the likelihood of the next character \cite{mott2017improving} or combining eye-gaze with other modalities such as humming \cite{hedeshy2021hummer}, head gestures \cite{feng2021hgaze}, foot-input \cite{rajanna2022presstapflick}, or touch \cite{kumar2020tagswipe}.
For entering text in smart glasses, eye-gaze has also been combined with touch \cite{ahn2019gaze} and in VR it has been shown that combining eye-gaze with controller selection works better than dwell \cite{rajanna2018gaze}. 
In addition, there have also been experiments on combining eye-gaze with brain-computer interfaces \cite{ma2018combining}.
Most recently, Hu et al. \cite{hu2024skimr} have presented a dwell-free eye-typing method that uses a statistical decoder to infer text from the eye-movement of the user. Another approach by Ren et al. \cite{ren2024eye} uses eye-gaze to assist finger input in augmented reality.
However, work combining eye-gaze and input through physical keyboards remains underexplored, even though key-presses have the potential to be an efficient confirmation method, specifically in a stationary setting when users sit in front of a physical keyboard. Hence, we have explored this combination in our work, which focuses specifically on entering accented characters in VR.

Apart from input modalities, prior work has also investigated various layouts for displaying the keyboard, such as circular \cite{yu2018pizzatext, jiang2020hipad} or cubic layouts \cite{yanagihara2018cubic}.
Also, the users' hands have been used to display the keyboard \cite{pratorius2014digitap, ogitani2018space, wang2015palmtype}.
In addition, it has been proposed to change the layout of physical keyboards, for example to avoid shoulder surfing while entering a password \cite{schneider2019reconviguration, maiti2017preventing} or to display a wide range of characters or even images \cite{schneider2019reconviguration}.

\subsection{Using Physical Keyboards in VR for Text Entry}
As typing on a physical keyboard still can be considered the gold standard in terms of efficient text entry, it has also been studied how to combine the physical device with VR.

McGill et al. \cite{mcgill2015dose} proposed to blend a video of the hands and keyboard into VR which resulted in increased performance while typing in VR.
Others have also tracked the hands and keyboard to visualize them in VR \cite{knierim2018physical, grubert2018effects} and different representations have been explored.
This has for example shown, that rendering only the fingertips can be as efficient as video-see-through \cite{grubert2018effects} and that partially transparent hands can be beneficial for inexperienced typists \cite{knierim2018physical}. 
Grubert et al. \cite{grubert2018text} have also shown that the position of the virtual keyboard can deviate from the physical without substantially influencing the typing performance and Pham et al. \cite{pham2019hawkey} presented a keyboard on a hawker's tray to allow typing while standing.
While several years ago spatial tracking systems were used to accurately track the keyboard and hands such as an OptiTrack system \cite{schneider2019reconviguration, knierim2018physical, grubert2018effects, grubert2018text}, current VR headsets (such as the Meta Quest 2, 3 or Pro) offer built-in tracking of both hands and keyboards.
Therefore, we are using the built-in capabilities of a Meta Quest Pro to track both the keyboard and hands which are then visualized through virtual representations in VR.

\subsection{Typing Special Characters}
Most prior work has focused on standard text entry in VR.
Yet, there has been some work on efficient special character entry in VR by recognizing the user's wrist orientation and changing the virtual keyboard accordingly \cite{song2022efficient}. 
Also, Wan et al. \cite{wan2024design} proposed techniques to enter both alphanumeric and special characters in VR by changing the content of keys while pointing at them.
Our work was mostly inspired by Schneider et al. \cite{schneider2019reconviguration} who proposed to press a modifier key in combination with a base letter to replace the original keys with a range of accented characters related to that letter, which can then be selected by pressing the corresponding key.
This can be very helpful when typing words from a different language or proper names that contain accented characters not present on the physical keyboard. As eye tracking is now available in the current VR headsets, we are combining the technique proposed by Schneider et al. \cite{schneider2019reconviguration} with eye-gaze, which could further enhance the input of accented characters. It can additionally help users keep their hands in the standard position on the keyboard, as it is not required to move them in order to press one of the accented characters.

Outside of VR, to enter such accented characters, users may switch the keyboard layout to a different language, which still requires referencing the layout to find the desired character. Alternatively, these characters can be entered by pressing the Alt key, entering a certain three-digit code for the respective character, and releasing the Alt key to complete the input.  
This inspired us to investigate a similar yet context-aware technique in VR, that alleviates the look-up of the related three-digit code.

\subsection{Multimodal Interaction Including Eye-Gaze}
As mentioned above, some previous work has already looked at multimodal text input in VR using eye-gaze \cite{ahn2019gaze, rajanna2018gaze, ma2018combining, ren2024eye}.
In other areas, eye-tracking has also been successfully combined with various modalities to support interaction in VR.
For example, eye-tracking has been combined with touch to interact with multiple displays \cite{biener2020breaking} or to enable one-handed input on a tablet \cite{pfeuffer2016gaze}.
Others have combined eye-tracking with gestures, for example, to select and manipulate objects \cite{pfeuffer2017gaze+} to select distant objects \cite{schweigert2019eyepointing}.
A survey of gaze-based interaction techniques was presented by Plopski et al. \cite{plopski2022eye}. These results encouraged us to further explore the combination of eye-tracking and other modalities for text entry in VR.

\section{Prototype Design}
\label{sec:prototype-design}
To facilitate special character input in VR on a physical keyboard, we present three different techniques which are described in the following:

\paragraph{\textsc{baseline}:} 
As mentioned before, for entering accented characters on a physical keyboard, users generally need to press Alt and a three-digit code that they need to remember or look up. 
For our VR technique, we enhanced this approach so that the users do not need to recall codes from memory. Upon pressing the modifier key Alt and a base character that is related to the desired accented character, relevant accented characters and their corresponding codes are displayed on the virtual monitor as long as the modifier key is being held down.

For example, in Fig. \ref{fig:teaser}, a) the user presses Alt and e to display accented characters related to e (è, é, ê, ë, ě).
Then, while holding down Alt, the user can enter the three-digit code and by releasing Alt the entry is confirmed and the accented character will be added to the text.

\paragraph{\textsc{reconviguration}:}
We adopted this technique from Schneider et al. \cite{schneider2019reconviguration}.  While following the original idea of using a modifier key to temporarily change key mappings, we change the layout of the accented keys relative to the base key. Specifically, upon pressing a modifier key (Alt) and a base character on the keyboard, the keys one row above the character that was pressed will change into different variations of that character as long as the modifier key is being held down.  Additionally, the variations are given a different base color to differentiate them from the regular keys as seen in Fig. \ref{fig:teaser}, b). Now the desired character can be selected by pressing the key corresponding to that character. This technique is inspired by the keyboard layout on touch devices such as smartphones.

\paragraph{\textsc{multimodal}:}
This technique combines keyboard input with eye-tracking.
The initial idea was to extend the \textsc{reconviguration} technique such that the selection of accented characters is performed through eye gaze to reduce hand and finger movement. Yet, with current hardware, the eye tracking was not reliable enough to select keys on the keyboard.
Therefore, upon pressing a modifier key (Alt) and a base character,
a list of square buttons each with a side-length of 6~cm, resulting in a vertical visual angle ranging from about 4.3° to 4.5° and a horizontal visual angle ranging from 4.6° to 4.8°, will be shown side by side at the bottom of the screen of the virtual monitor. 
\newmarker{As the size of all squares is the same, the visual angle varies slightly depending on their position relative to the user (squares in the center have a greater visual angle).}
They are shown as long as the modifier key is being held down. 
\newmarker{The size of the squares was determined empirically through informal testing with 5 people who agreed on a size that allows for an accurate selection of the individual buttons given the Meta Quest Pro eye tracker.}
Now the desired character can be targeted via eye gaze and it will change its base color to indicate targeting. Once the desired character has been targeted, it can be selected by pressing the key that was used to initialize the action, i.e., the base key once again. When an accented character is entered successfully, the background of the targeted key will change to a green color for a short duration (100ms) to provide the user with visual feedback, see Fig. \ref{fig:teaser}, c).

\section{Implementation}

The prototype was implemented using Unity version 2020.3.33f1. The VR headset used for the study was the Meta Quest Pro, which provided the required eye-tracking and hand-tracking as well as the tracking of various physical keyboards. 
Among such keyboards, we used the Logitech MX Keys S, as it has a numeric keypad (numpad), which is necessary to enter numeric codes efficiently.
For the study, we used an English (US) keyboard layout.

For the virtual environment, we created a scene using the Space Setup function on the VR headset itself. We allowed our application access to the scene to determine the position and geometry of the room (floor, ceiling, walls) and the table, where the keyboard was positioned. Following ergonomic guidelines \cite{long2014visual}, we placed a virtual monitor on the table inside the virtual environment at a distance of 65~cm from the user, with the top of the screen at eye level. 
We additionally implemented functionality to adjust the height of the monitor or to move it further left or right by using the arrow keys of the keyboard while holding down the Ctrl key to allow further individualization by the users.

An input field was placed at the center of the virtual monitor. A text box displaying the instructions or the actual stimulus that the user was supposed to type, was placed above the input field (see Fig, \ref{fig:teaser}), providing a vertical visual angle of around 4.1 degrees per line. The space underneath the input box was reserved to show additional information depending on the current input technique. 
In \textsc{baseline}, we showed a table of two rows, where the first row showed the accented characters related to the base character, and the second row showed their corresponding numeric codes.
In \textsc{reconviguration}, this space was unused, and in \textsc{multimodal}, we populated this space with square buttons, displaying the related accented characters for selection via eye-gaze.






\section{User Study}
Our user study investigated the performance differences between the three proposed techniques (c.f. Sect. \ref{sec:prototype-design}) to support accented character entry on physical keyboards, in VR. To this end, we designed an experiment in which participants had to type a list of names consisting of accented characters, using the provided keyboard with US English layout.

This user study was conducted as a 1$\times$3 within-subjects design with the independent variable \textsc{technique}, with three levels, \textsc{baseline}, \textsc{reconviguration}, and \textsc{multimodal}, resulting in three conditions of the user study. The order of the conditions was balanced. 
For the objective dependent variables, we recorded the text entry speed in terms of words per minute (WPM) and accuracy in terms of character error rate (CER). Both text entry speed and accuracy were measured for overall text, accented characters, and regular English characters. 

We additionally collected subjective data including the usability (system usability scale (SUS) \cite{brooke1996sus}), task load (Raw NASA TLX questionnaire (TLX) \cite{hart1988development}), simulator sickness (simulator sickness questionnaire (SSQ) \cite{kennedy1993simulator}) and a preference questionnaire. 
Further, we conducted semi-structured interviews after each condition and at the end of the user study. After each condition, we asked the participants to report on their likes and dislikes of the respective technique. After all conditions, we asked them to report on their favorite technique, if they could imagine using it in the future, 
and for any recommendations on improvements.


\subsection{Task}
The participants had to perform a typing task during the experiment. As our prototype focused specifically on entering accented characters from various languages, we did not use sentences as a stimulus for text entry. Instead, we used word lists as stimuli, where each word of the word lists consisted of accented characters. We chose the accented characters from French, German, Danish, and Turkish, that were related to vowels in English.
Table \ref{tab:accented-characters} shows the chosen accented characters.
We chose the stimuli words from human first names as well as city names. These were acquired by scraping Wikipedia \cite{wiki:given_names}  for human first names and by processing a dataset of all cities with a population of over 500 from opendatasoft \cite{opendatasoft} for the city names. Filtering the data by the chosen accented characters, gave us a total of 597 first names and 20,876 city names. To reduce variation in word length we decided to only include words with a length of five to eight characters (including spaces). This left us with 431 human first names and 7,825 city names from which we generated the word lists.

We then sampled 1560 names from the set of 8256 words that were five to eight-characters long. We then separated the sampled words into six lists, two for each condition with one for training and the other for the actual task. The training lists and the task lists had 120 words and 400 words each, respectively. We counterbalanced the lists used with the order of conditions. The order within the task lists was randomized during the experiment. However, the training list was fixed, and the first 29 words contained all different accented characters to ensure that the participants had the opportunity to practice all accented characters during training.
During the experiment, the participants were shown the stimuli words one after the other, on the virtual screen. Their task was to type the same word in the input text box on the virtual screen, as fast as possible.

\begin{table}[h]
    \centering
    \caption{Accented characters used in the user study.} 
    \begin{tabular}{|c|c|}
        \hline
        Base Characters & Accented Characters \\
        \hline
        a & á, à, â, ã, ä, å, æ \\
        A & Á, À, Â, Ã, Ä, Å, Æ  \\ \hline
        e & è, é, ê, ë, ě \\
        E & È, É, Ê, Ë, Ě \\ \hline
        i & ì, í, î, ï, ı \\
        I & Ì, Í, Î, Ï, İ \\ \hline
        o & ò, ó, ô, õ, ö, ø, œ \\
        O & Ò, Ó, Ô, Õ, Ö, Ø, Œ \\ \hline
        u & ù, ú, û, ü, ů \\
        U & Ù, Ú, Û, Ü, Ů \\
        \hline
    \end{tabular}
    \label{tab:accented-characters}
\end{table}




\subsection{Participants}
We recruited nineteen participants in the study. Due to technical problems, we lost the data of the first participant (P01). Eighteen participants (P02-P19) completed the study (12 male, 6 female, mean age 25.83 years, sd = 4.6) and hence the study design was overall balanced.
Participants had to be at least 18 years old to participate in the study. Six participants wore glasses during the user study.
Nine of the participants usually typed in English, and six typed in German. One participant typed both in English and German, one in Russian and German, and one in English and Persian. Twelve participants were university students, one designer, two researchers, and three software developers. We additionally asked the participants how often they looked at the keyboard while typing and about their experience with VR, which is summarized in Table \ref{tab:demographics}.

\begin{table}[h]
    \centering
    \small
    \vspace{0cm}
    \caption{Number of participants with respective touch typing skills and VR experience.} 
    \begin{tabular}{|c|c|c|c|c|c|}
        \hline
        \cline{2-6}
             &\multicolumn{1}{|c|}{Never} 
             &\multicolumn{1}{|c|}{A Little}
             &\multicolumn{1}{|c|}{Moderate}
             &\multicolumn{1}{|c|}{Often}
             &\multicolumn{1}{|c|}{Very Much}
         \\
        \hline
            Touch Typing & 0 & 7 & 11 & 0 & 0 \\ \hline 
            VR Experience & 4 & 5 & 6 & 1 & 2 \\ 
        \hline
    \end{tabular}
    \label{tab:demographics}
\end{table}


\subsection{Procedure}
\begin{figure}[t]
	\centering
	\includegraphics[width=\linewidth]{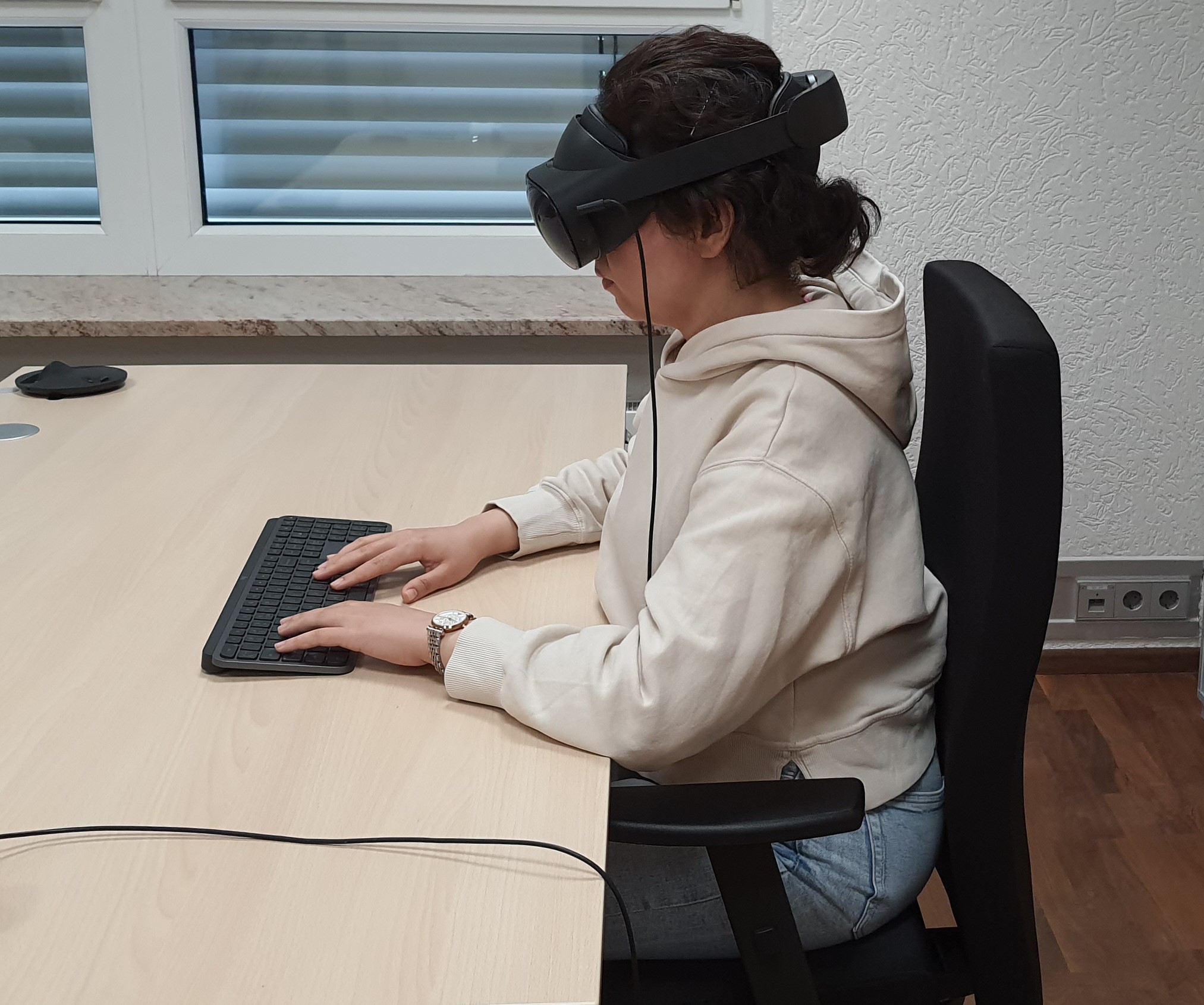}
        \vspace{0cm}
	\caption{Setup for the user study: Participant was wearing the Meta Quest Pro with the Logitech MX Keys S Keyboard in front of them.
        }
	\label{fig:study-setup}
\end{figure}

All participants were informed about the procedure of the study. They signed a consent form, filled out a demographic questionnaire and the experiment started. The study setup is shown in Fig. \ref{fig:study-setup}.
At the beginning of the experiment, we showed the participants all the accented characters and their related English characters, which were used in the study.
Before each condition, we explained to them the respective text entry technique orally.
After putting the VR headset on, the participants could adjust the position and height of the virtual screen, as per their comfort.

The participants started the first condition with a short training session of 3 minutes. After this, they performed the typing task, which lasted for 10 minutes.
After finishing, they answered the system usability scale questionnaire, the Nasa TLX, and the simulator sickness questionnaire, after which a short interview of 2 minutes was conducted.
This procedure was repeated for the remaining two conditions.
We performed eye-gaze calibration using the Meta Quest Pro calibration routine right before the \textsc{multimodal} condition.
Finally, the participants filled out the preference questionnaire and ranked the techniques. Lastly, an interview was conducted to reflect on the differences between the conditions.
The overall duration of the user study was 90 minutes and each participant was compensated with a 20€ gift card for their time. 

\subsection{Results}

The influence of \textsc{technique} was analyzed using repeated measures analysis of variance (RM-ANOVA). 
For the non-normal objective data, we first checked for sphericity. When the sphericity was met, we used RM-ANOVA, as normality violations do not tend to have a major impact on the robustness of the analysis \cite{blanca2023non}. When the sphericity was not met, we used Greenhouse-Geisser correction prior to RM-ANOVA.
We used Aligned Rank Transform ~\cite{wobbrock2011aligned} for subjective data that were not normally distributed.
Bonferroni adjustments were used for multiple comparisons at an initial significance level of $\alpha=0.05$.

The descriptive statistics are presented in Table \ref{tab:descriptiveStatistics}. The results of the statistical tests are presented in Table \ref{tab:resultsTable}.

\begin{table}[h]
    \centering
    \small
    \caption{Descriptive statistics of the dependent variables \newmarker{including the mean ($m$) and standard deviation ($sd$)}. BL=\textsc{baseline}, RVR=\textsc{reconviguration}, MM=\textsc{multimodal}. Measures with significant findings are marked in gray.} 
    \begin{tabular}{|c|c|c|c|}
        \hline
            
             \multicolumn{1}{|c|}{ \textbf{Dependent Variables}}
             &\multicolumn{1}{|c|}{ \textbf{BL ~$m~(sd)$}} 
             &\multicolumn{1}{|c|}{ \textbf{RVR ~$m~(sd)$}}
             &\multicolumn{1}{|c|}{ \textbf{MM ~$m~(sd)$}}
         \\
        \hline
            \rowcolor{lightgray}
            Overall WPM             & 7.23 (2.06)   & 9.66 (2.65)   & 9.05 (3.81) \\ \hline 
            \rowcolor{lightgray}
            Accented Character WPM  & 2.93 (1.11)     & 5.06 (1.19)   & 5.17 (2.27) \\ \hline
            Regular Character WPM   & 24 (7.83)     & 25.9 (7.97)   & 25 (10.01) \\ \hline
            Overall CER             & 0.013 (0.014) & 0.015 (0.011) & 0.009 (0.007) \\ \hline
            Accented Character CER  & 0.035 (0.04) & 0.049 (0.044)  & 0.041 (0.036)  \\ \hline
            Regular Character CER   & 0.014 (0.019) & 0.016 (0.013) & 0.01 (0.010) \\ \hline
            \rowcolor{lightgray}
            System Usability        & 54.4 (23.01)  & 77.36 (14.64) & 68.05 (16.99) \\ \hline
            Task Load               & 45.98 (18.37) & 37.89 (12.41) & 43.33 (14.38) \\ \hline
            Simulator Sickness      & 17.03 (18.77) & 17.87 (20.94) & 19.32 (20.97) \\
        \hline
    \end{tabular}
    \label{tab:descriptiveStatistics}
\end{table}

\begin{table*}[t]
    \centering 
    \caption{RM-ANOVA results for the user study. d$f_1$ = d$f_{effect}$ and d$f_2$ = d$f_{error}$. Significant findings are marked in gray.
    }
    \setlength{\tabcolsep}{5pt}
        
        \begin{tabular}{|c||c|c|c|c|c||c|c|c|c|c||c|c|c|c|c|}
            \hline
            &\multicolumn{5}{|c|}{Overall WPM}
            &\multicolumn{5}{|c|}{Accented Character WPM} 
             &\multicolumn{5}{|c|}{Regular Character WPM}

            \\
            \cline{2-16}
            & d$f_1$ & d$f_2$ & F & p &  $\eta^2_p$
            & d$f_1$ & d$f_2$ & F & p &  $\eta^2_p$ 
            & d$f_1$ & d$f_2$ & F & p &  $\eta^2_p$ 
            \\
            \hline
            \textsc{technique}  &$2$ & $34$ & $8.61$ & \cellcolor{lightgray}$<.001$ & $0.336$
                                & $2$ & $34$ & $16.7$ & \cellcolor{lightgray}$<0.001$ & $0.495$ 
                                & $2$ & $34$ & $1.82$ & $0.177$ & $0.097$
                                 \\             
            \hline                    
       \end{tabular}
        \\ \vspace{0 cm}
        \begin{tabular}{|c||c|c|c|c|c||c|c|c|c|c||c|c|c|c|c|}
            \hline
            &\multicolumn{5}{|c|}{Overall CER} 
            &\multicolumn{5}{|c|}{Accented CER}
            &\multicolumn{5}{|c|}{Regular CER} 
            \\
            \cline{2-16}
            & d$f_1$ & d$f_2$ & F & p &  $\eta^2_p$
            & d$f_1$ & d$f_2$ & F & p &  $\eta^2_p$
            & d$f_1$ & d$f_2$ & F & p &  $\eta^2_p$
            \\
            \hline
            \textsc{technique}  &$2$ & $34$ & $2.00$ & $0.151$ & $0.105$
                                & $2$ & $34$ & $0.691$ & $0.508$ & $0.039$ 
                                & $1.52$ & $25.79$ & $1.16$ & $0.317$ & $0.064$
                                \\             
            \hline                    
        \end{tabular}
        \\ \vspace{0 cm}
         \begin{tabular}{|c||c|c|c|c|c||c|c|c|c|c||c|c|c|c|c|}
            \hline
            &\multicolumn{5}{|c|}{System Usability} 
            &\multicolumn{5}{|c|}{Task Load} 
            &\multicolumn{5}{|c|}{Simulator Sickness}
            \\
            \cline{2-16}
            & d$f_1$ & d$f_2$ & F & p &  $\eta^2_p$
            & d$f_1$ & d$f_2$ & F & p &  $\eta^2_p$ 
            & d$f_1$ & d$f_2$ & F & p &  $\eta^2_p$
            \\
            \hline
            \textsc{technique}  &$2$ & $34$ & $5.874$ & \cellcolor{lightgray}$0.006$ & $0.257$
                                & $2$ & $32$ & $2.683$ & $0.084$ & $0.143$ 
                                & $2$ & $34$ & $0.494$ & $0.614$ & $0.028$
                                 \\             
            \hline                    
        \end{tabular}
        \label{tab:resultsTable}
\end{table*}

\begin{figure*}[t]
	\centering 
	\includegraphics[width=\linewidth]{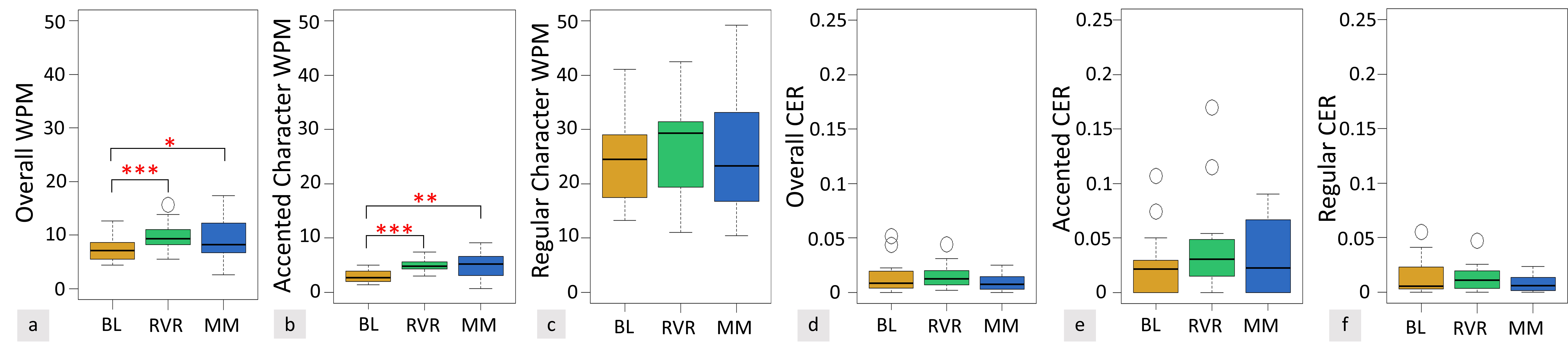}
	\caption{
         Box plots for objective dependent variables: a) Overall WPM. b) Accented Character WPM. c) Regular Character WPM. d) Overall CER. e) Accented CER. f) Regular CER. BL=\textsc{baseline}, RVR=\textsc{reconviguration}, MM=\textsc{multimodal}. 
         The number of stars indicates the significance levels between the techniques: *** \textless 0.001, **  \textless 0.01, *  \textless 0.05. 
        }
	\label{fig:wpm-cer-plots}
\end{figure*}

\subsubsection{Text Entry Rate}
The text entry speed was calculated in terms of WPM. We computed it for all characters, i.e., accented and regular, specifically for accented characters, and lastly for regular characters only.  Fig. \ref{fig:wpm-cer-plots} shows the box plots for text entry speed.

\paragraph{Overall WPM.} We define this metric as the length of the stimulus divided by the length of a word, defined as 5 characters including spaces, divided by the time it took the user to complete the entry.

We found a significant effect of \textsc{technique} on the overall WPM \newmarker{($F(2,34) = 8.61, ~p < 0.001, ~\eta^2_p = 0.336$)}. 
Posthoc tests revealed a significantly higher overall text entry speed with \textsc{reconviguration} compared to \textsc{baseline} ($ p < 0.001$, Cohen's $d_z = 1.44 $).
It was also significantly higher with \textsc{multimodal} compared to \textsc{baseline} ($ p = 0.038$, Cohen's $d_z = 0.655 $), \newmarker{as shown in Fig. \ref{fig:wpm-cer-plots}(a)}. 
But, we did not find a significant difference between \textsc{reconviguration} and \textsc{multimodal} ($ p = 1.00$, Cohen's $d_z = 0.199 $).
The overall text entry speed was 33.6\% higher with \textsc{reconviguration} as compared to \textsc{baseline} and was 25.17\% higher with \textsc{multimodal} as compared to \textsc{baseline}.

\paragraph{Accented Character WPM.} This metric was calculated only on the correct inputs as to avoid inaccuracies caused by unintended mechanisms such as key repeat and to have an accurate measurement solely based on the accented characters. On average 92.9\% (sd = 5.4\%) of inputs were found to be correct.
For each input, the time it took to enter each accented character was calculated by taking the difference between the time when the accented character was entered and the time since the previous operation. An operation is defined as a deletion or an insertion of a character. 
Then, the number of accented characters is divided by the word length of 5 characters and the result is divided by the sum of the individual times. Finally, the mean is calculated.

We found a significant effect of \textsc{technique} on the accented character WPM \newmarker{($F(2,34) = 16.7, ~p < 0.001, ~\eta^2_p = 0.495$)}.  
Posthoc tests revealed a significantly higher accented character entry speed with \textsc{reconviguration} compared to \textsc{baseline} ($ p < 0.001$, Cohen's $d_z = 2.249 $). 
It was also significantly higher with \textsc{multimodal} compared to \textsc{baseline} ($ p = 0.003$, Cohen's $d_z = 0.925 $), \newmarker{as shown in Fig. \ref{fig:wpm-cer-plots}(b)}. 
But, we did not find a significant difference between \textsc{reconviguration} and \textsc{multimodal} ($ p = 1.00$, Cohen's $d_z = -0.055 $).
The accented character text entry speed was 72.7\% higher with \textsc{reconviguration} as compared to \textsc{baseline} and was 76.4\% higher with \textsc{multimodal} as compared to \textsc{baseline}.

\paragraph{Regular Character WPM.} Similar to the accented character WPM, this metric was also calculated using only the correct inputs. For each input, the time it took to enter each non-accented character was calculated by taking the difference between the time when the non-accented character was entered and the time since the previous operation. An operation is defined as a deletion or an insertion of a character. Then the number of non-accented characters are divided by the word length of 5 characters and the result is divided by the sum of the individual times. Finally, the mean is calculated.

The interface for entering regular English characters was the same in all the techniques, hence, as expected, we did not find a significant effect of \textsc{technique} on the regular character entry speed, \newmarker{as shown in Fig. \ref{fig:wpm-cer-plots}(c)}.

\subsubsection{Accuracy} 
Similar to the text entry speed, we computed the accuracy for all characters, i.e., accented and regular, specifically for accented characters, and lastly for regular characters only. Fig. \ref{fig:wpm-cer-plots} shows the box plots for accuracy.

\paragraph{Overall CER.} This metric is the minimum number of character-level insertion, deletion, and substitution operations required to transform the user input into the stimulus text, divided by the number of characters in the stimulus text. \textsc{technique} did not have a significant effect on the overall CER, \newmarker{as shown in Fig. \ref{fig:wpm-cer-plots}(d)}.

\paragraph{Accented CER.} First, for each input the position of the accented characters from the stimulus text was compared with the characters at the same position in the user input. The number of mismatches was then divided by the number of accented characters in the stimulus text yielding the resulting metric. \textsc{technique} did not have a significant effect on the accented CER, \newmarker{as is shown in Fig. \ref{fig:wpm-cer-plots}(e)}.

\paragraph{Regular CER.} This metric was calculated by first deleting the characters at the positions of the accented characters from the stimulus text as well as user input.
After deletion, the number of character-level insertion, deletion, and substitution operations required to transform the input text to the stimulus text was divided by the number of characters in the stimulus text to yield the final result.
\textsc{technique} did not have a significant effect on the regular CER, \newmarker{as shown in Fig. \ref{fig:wpm-cer-plots}(f)}.


\subsubsection{Subjective Measures}

\begin{figure}[t]
	\centering 
	\includegraphics[width=\linewidth]{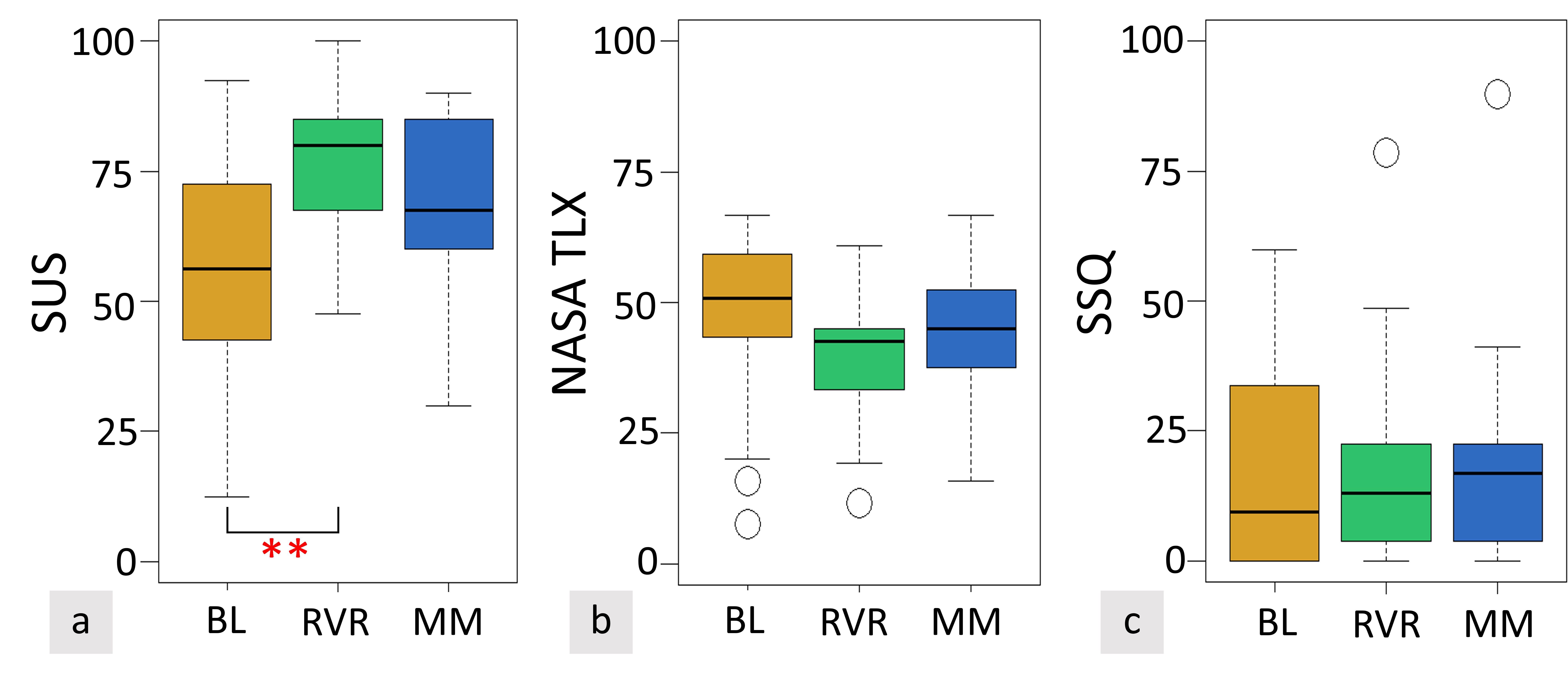}
	\caption{
        Box plots for subjective measures: a) System Usability Scale Scores. b) NASA Task Load Index. c) Simulator Sickness. BL=\textsc{baseline}, RVR=\textsc{reconviguration}, MM=\textsc{multimodal}. 
         The number of stars indicates the significance levels between the techniques: *** \textless 0.001, **  \textless 0.01, *  \textless 0.05. 
        }
	\label{fig:questionnaire-plots}
\end{figure}

A significant effect of \textsc{technique} was found on the system usability \newmarker{($F(2,34) = 5.874, ~p = 0.006, ~\eta^2_p = 0.257$)}. 
Posthoc tests revealed significantly higher ($ p = 0.005$, Cohen's $d_z = 0.805$) usability of \textsc{reconviguration} compared to \textsc{baseline}, \newmarker{as shown in Fig. \ref{fig:questionnaire-plots}(a)}. 
\textsc{reconviguration} had a 42\% higher usability than \textsc{baseline}. 
No significant differences were found in the pairs \textsc{multimodal} and \textsc{baseline} ($ p = 0.183$, Cohen's $d_z = 0.457$) or \textsc{reconviguration} and \textsc{multimodal} ($ p = 0.445$, Cohen's $d_z = 0.348$). 

The independent variable \textsc{technique} did not have a significant effect on either the overall task load or overall simulator sickness\newmarker{, see Fig. \ref{fig:questionnaire-plots}(b) and \ref{fig:questionnaire-plots}(c),} nor on any of their respective sub-scales. 
\deletemarker{Fig. 4 shows the box plots for all the subjective measures.}

\subsubsection{User Feedback}
The responses from the preference questionnaire indicated that 44.4\% of the participants preferred \textsc{reconviguration}, 44.4\% preferred \textsc{multimodal }, and 11.1\% preferred \textsc{baseline}.

Participant who preferred \textsc{reconviguration} mentioned that it is easy to use (P03, P05, P18, P19), feels natural (P08, P17), and is fast (P05, P17). P03 also mentioned that she had to adapt to the technique but learned it fast. P08 mentioned that it is close to what she knows from using a smartphone and P12 and P16 liked that the accented characters were displayed on the keyboard where they are already looking while typing.
The participants who preferred \textsc{multimodal} reported that it was fun (P06, P14) and intuitive (P06, P09). P14 perceived it as faster, P10 as easy and P02 found it very exciting.
Three participants liked that they do not need to look down at the keyboard (P07, P13, P15) and P14 noticed that it requires less head movement.
One participant who preferred \textsc{baseline} found it challenging to learn the letters and compared it to a game (P04). 
The other one (P11) said he rarely looks at the keyboard so this technique does not break his routine.

In addition, after each condition participants were asked what they liked and disliked about the respective condition. 
Regarding the \textsc{baseline} technique, participants found it fun (P02, P04, P05, P12), interesting (P02) and easy to use (P13, P18). 
P17 mentioned it is better than the Windows default system and P18 suggested it could also be used without VR. 
P05 mentioned it worked better than \textsc{multimodal} and P06 said the use of Alt and numpad was muscle memory friendly.
However, four participants had problems finding the buttons on the numpad (P01, P07, P08, P10).
P12 found the technique highly complex and others mentioned it took lots of efforts (P16), was slow (P03), interrupted the flow when moving to the numpad (P06) and was not good for touch-typists (P09).

Regarding the \textsc{reconviguration} technique, participants found it easy to use (P03, P04, P08, P09, P12) and fast (P02, P03, P05, P14).
Four participants emphasized that they liked it (P13, P14, P18, P19). P08 said she got used to it fast and P05 mentioned she did not have to think a lot.
P06 found the design clever, P10 liked the availability of the accented characters in front of the eyes and found it more comfortable than the other conditions and for P07 it was a very new experience.
On the other hand, three participants did not like to look down at the keyboard (P06, P07, P18) and P04 found it difficult to reach the buttons.

Regarding the \textsc{multimodal} technique, seven participants said they liked it a lot (P05, P06, P08, P09, P14, P16, P19). 
P03 felt like it was faster than other techniques and P04 said using less hand typing was a good idea.
P06 found it was fun and P09 was fascinated.
Three participants mentioned that it was intuitive (P06, P07, P12), three mentioned it is easy to learn (P03, P07, P10) and three said it is good that they don't need to look down (P07, P13, P16).
For three participants the eye tracker worked very well (P15, P17, P18).
However, for others another three the eye tracking did not work well (P02, P11, P14).
Three people also found the blinking annoying which resulted from highlighting the currently looked-at button (P04, P13, P16).
P12 and P15 mentioned they were not used to such a technique, P03 got tired and P04 complained about needing to focus a lot.

Participants were also asked to suggest improvements to the techniques.
For the \textsc{reconviguration} technique, P17 proposed to display the accented characters to the left or right of the pressed key, depending on the location of the base-character. P18 on the other hand would prefer if they were always displayed at a central location. P19 would prefer a larger font-size of the characters on the keys.
Regarding the \textsc{multimodal} technique, P06 suggested giving the user the ability to change the location of the presented characters. P19 mentioned that he did not like to press the key twice (once for selecting base-character and once for confirming eye-selection) and P17 suggested solving this by selecting the character by releasing the key.
For the \textsc{baseline} technique, P03 would prefer more systematic codes and P11 would like to choose the characters with the mouse or the arrow keys instead of typing the code.

When asked if they would use such techniques in the future P06 mentioned it would be good for people who type in multiple languages and P02 said using VR reduced the need for multiple keyboards for different languages. However, P10 also mentioned that the presented functionality could be replaced by AI which replaces the base-characters automatically.


\section{Discussion}
The presented study indicated that both the overall text entry speed and the accented character text entry speed were higher when using the \textsc{reconviguration} or \textsc{multimodal} techniques than when using the \textsc{baseline} technique.
\newmarker{This can be expected, as the \textsc{baseline} techniques require two more key presses per accented character than the other two techniques and can therefore be explained by the keystroke level model \cite{card1980keystroke} }.
However, these results illustrate how the capabilities of VR, such as eye-tracking or changing the layout of the physical keyboard, can be used to make special character entry more efficient.

We also found that the usability of \textsc{reconviguration} was rated significantly higher than that of \textsc{baseline}. 
The majority of participants also preferred either \textsc{reconviguration} or \textsc{multimodal} (44.4\% each).
From the interviews, we could infer that participants who usually look at the keyboard while typing prefer to select the accented characters on the keyboard and therefore like \textsc{reconviguration} the best. On the other hand, participants who generally look at the screen while typing preferred \textsc{multimodal} as it displays the selection of accented characters there. The strength of these VR implementations is, that both options could be made available to users, and, hence, allow for customization based on users' preferences (of course, only if eye-tracking is available). 
Another possibility would be to display a virtual copy of the keyboard close to the screen to make the \textsc{reconviguration} technique more attractive for touch-typists, as it has been found that repositioning the representation of a physical keyboard in VR does not negatively affect performance \cite{grubert2018text}. 

For all techniques, participants suggested improvements in the area of personalization, such as choosing the location in which the selection of the accented characters is displayed.
Such an option could be especially useful for non-touch-typists who need to look back and forth between the keyboard and screen when using the \textsc{multimodal} technique.
\newmarker{This could be alleviated by additionally performing eye selection on the keyboard, or close to the keyboard, as it has been demonstrated that gaze typing on the keyboard is viable, given that the entire keyboard is within view \cite{rajanna2018gaze}. However, for interaction with the small keys on a physical keyboard, eye tracking accuracy will need to increase for robust selection.}

Therefore, we would recommend future systems to allow users to adjust such settings according to their personal typing preferences. This is also very feasible in VR as there are no physical restrictions.
All three techniques could also be transferred to an augmented reality (AR) setting, as it is also possible to replace characters on a physical keyboard in AR \cite{maiti2017preventing}, yet there might be some restrictions on the placement of the accented characters.
Also, depending on the users' preferences it could be beneficial to offer multiple techniques such as the \textsc{reconviguration} and \textsc{multimodal} techniques.
\newmarker{Future work should explore investigating adaptive techniques that adjust user interaction based on user behavior and preferences to help cater to individual needs and optimize usability.}

For future work, it would also be interesting to explore further variations of the presented techniques. 
For example, the \textsc{baseline} technique could allow the selection of accented characters through a mouse-click or the arrow-keys to switch between the presented characters, as proposed by P11.
Also, as proposed by P17, accented characters could be selected by key-release instead of pressing the key twice when using the \textsc{multimodal} technique.
Another option for both \textsc{reconviguration} and \textsc{multimodal} would be to already show the list of accented characters before pressing the base character while only touching it or even only hovering above it.
\newmarker{The proposed techniques could also be combined with artificial intelligence, for example, to correct incorrectly accented characters caused by inaccurate eye tracking.}


    



\section{Limitations}
One limitation of the presented study is the stimulus selection. We only used first names and city names and the base characters of all accented characters that we used in the study were vowels. In regular text entry tasks, researchers have developed established phrase sets~\cite{mackenzie2003phrase}, which have been refined to capture mobile text~\cite{vertanen2011versatile}, children's text~\cite{kano2006children}, and additional language traits~\cite{paek2011sampling}. The memorability and presentation style has also been demonstrated to affect performance results in regular text entry tasks~\cite{kristensson2012performance}. It would be useful to standardize the task for this area of text entry as well.

In addition, we only evaluated the performance after a short training period.
From the user feedback, we learned that \textsc{reconviguration} could be considered to be rather easy to learn and use.
To be proficient with the \textsc{multimodal} technique, on the other hand, might take longer, as most participants are not very experienced with eye-tracking input in the first place. This is also indicated by the mean accented character speed in  \textsc{multimodal} ($m=5.17$), which is slightly higher than \textsc{reconviguration} ($m=5.06$).
Therefore, it would be insightful to repeat the study after a longer practice period, \newmarker{as long-term studies will help in understanding the learning curve of each technique and hence the sustained performance.}

It is also possible that the change in mode from key input to eye input and then again key input in the \textsc{multimodal} technique could cause a lower typing speed.
This problem is most likely intensified for non-touch-typists who need to look back and forth between the keyboard and screen.
As mentioned before, this could be alleviated by performing the eye-selection closer to the keyboard.
\newmarker{Further, it would be important to investigate the effects of prolonged use of eye tracking interfaces.}

\newmarker{Next, while prototype designing, the size of the squares displaying the characters in the \textsc{multimodal} technique was based on the subjective opinion of a small sample. Further research is encouraged to find an optimal size quantitatively with a larger sample, potentially further enhancing the performance of this technique. However, finding a universally valid size might be difficult, as eye tracking quality may vary across different VR headsets and different positions of the display within the same VR headset, which would likely influence the required size. Hence, we advise for more thorough future investigations of eye-tracking accuracy across eye-tracking capable VR headsets.}

Further, for \textsc{baseline}, we were inspired by a Microsoft Windows technique for accented character entry and only investigated the performance of a similar context-aware technique in our user study. Other techniques, e.g., the Compose Key technique from Linux should still be evaluated, in future work.

\newmarker{Finally, for the initial evaluation of our prototype we analyzed the data of 18 participants with experience in four different languages. In future studies, participants from a more diverse group, including more female participants, a broader age group and further linguistic backgrounds, should be included to increase generalizability.}


\section{Conclusions and Future Work}
In this paper we yave presented three techniques for entering accented characters in VR using a physical keyboard. 
First, the \textsc{baseline} technique which presents the user with codes of relevant accented characters that then need to be entered on the numpad.
Second, the \textsc{reconviguration} technique presented by Schneider et al. \cite{schneider2019reconviguration} which changes the keyboard layout to display relevant accented characters which can then be selected via key-press.
Third, a \textsc{multimodal} technique which presents the user with relevant accented characters that can be selected via eye-gaze and confirmed through a key-press.
We found that both the \textsc{reconviguration} and \textsc{multimodal} techniques resulted in significantly faster text entry speed than \textsc{baseline} and they were each preferred by 44.4\% of participants. They were also both found to be more usable than \textsc{baseline}, yet only significantly for \textsc{reconviguration}.
From participants' feedback, we could infer that \textsc{multimodal} was more appropriate for touch-typists, as they did not want to look at the keyboard, while \textsc{reconviguration} was preferred by people who regularly look at the keyboard. Therefore, in combination with other suggestions from participants, we recommend to make future systems highly personalizable to satisfy various user needs.


\bibliographystyle{abbrv-doi}

\balance
\bibliography{template}
\end{document}